\documentclass[aps,prc,twocolumn,showkeys,showpacs,fleqn]{revtex4}
\usepackage{amssymb}
\usepackage{graphicx}% Include figure files
\usepackage{color}

\usepackage{amsmath}
\newcommand{\req}[1]{(\ref{#1})}
\newcommand{\bel}[1] { \begin{equation} \label{#1}} %@#1
\newcommand{\belar}[1] { \begin{eqnarray} \label{#1}} %@#1
\newcommand{\ee} { \end{equation}} %@#1

\begin{document}
\title{Description of the mass-asymmetric fission of the Pt isotopes,
obtained in the reaction $^{36}$Ar + $^{142}$Nd within the two-stage fusion-fission model}
\author{V. L. Litnevsky}
\email{vlad.lit@bk.ru}
\affiliation{Omsk state transport university, Omsk, Russia}

\author{ F. A. Ivanyuk}
\email{ivanyuk@kinr.kiev.ua}
\affiliation{Institute for Nuclear Research, Kiev, Ukraine}

\author{G. I. Kosenko}
\email{kosenkophys@gmail.com}
\affiliation{Omsk Tank Automotive Engineering Institute, Omsk, Russia}

\author{S. Chiba}
\email{chiba.satoshi@nr.titech.ac.jp}
\affiliation{Tokyo Institute of Technology, Tokyo, Japan}

\date{today}
\begin{abstract}
The two stages dynamical stochastic model developed earlier for description of fusion-fission reactions is applied to the calculation of mass- and energy-distributions of fission fragments of platinum isotopes in reaction ${\rm ^{36}Ar + ^{142}Nd \to ^{178-x}Pt + xn}$. The first stage of this model is the  calculation of the approaching of projectile nucleus to the target nucleus. On the second stage of the model, the evolution of the system formed after the touching of the projectile and target nuclei is considered. The evolution of the system on both stages  is described by three-dimensional  Langevin equations for the shape parameters of the system.  The mutual orientation of the colliding ions and tunneling through the Coulomb barrier in the entrance channel are also taken into account. The potential energy of the system is calculated within the macroscopic-microscopic approach. The calculated mass-energy distributions of fission fragments are compared with the available experimental data. The impact of shell effects, rotation of the system and neutron evaporation on the calculated results is discussed.
\end{abstract}

\pacs{25.70.Jj,24.10.-i,21.60.Cs} \keywords{fusion-fission}

\maketitle

\section{Introduction}
The first theoretical description of nuclear fission discovered in 1939 \cite{Gan:39,meitner} was given in the framework of the liquid drop model \cite{Bohr:39}, in which it is assumed that the properties of the atomic nucleus are similar to those of a charged viscous incompressible liquid drop. It turned out that the liquid drop model predicts splitting of fissioning nucleus into two fragments with the same masses, what was confirmed by experiments of highly excited nuclei. However, in the case of low excitation energies, such as the fission of $^{235}$U by thermal neutrons, the masses of the fission fragments were not equal. The explanation of mass-asymmetric fission was given later by the influence of shell effects. The presence of mass-asymmetric fission valley for transuranium elements was demonstrated by Pashkevich in  \cite{Pashkevich:71}. In this work the dependence of the potential energy of nuclei on their deformation was constructed using the shell correction method of Strutinsky \cite{Strutinsky:67,Brack:72}
and it was shown that the fission process is strongly affected by the deformation dependence of the potential energy of the system. Further progress in the description of the process of nuclear fission is associated with the use of statistical \cite{Fong:69,mebel:92} and dynamical \cite{Adeev:89,usang2019} models, as well as their combinations \cite{Mavlitov:92}. These models describe the fission of a nucleus initially located in a potential well near the ground state (compound nucleus). Such an excited compound nucleus can be formed by irradiating of atomic nuclei by light particles.

The description of the fission process becomes much more complicated if the excited nucleus was formed in the result of fusion reaction of two massive ions. In this case, one can't talk merely about the fission of compound nucleus. From the moment of touching of the initial nuclei to the moment of formation of the compound nucleus passes quite a long time, during which the system may undergo fission or may reduce its excitation energy by emitting a light particle or gamma quanta and form the compound nucleus in the ground state. As the result, for the description the mass distribution of fission fragments it is necessary to consider the whole evolution of fusion-fission process starting from the approaching of the initial nuclei to each other and ending with the formation of compound nucleus or fission.

For this purpose in \cite{GI:02,Litnevsky:14_1, Litnevsky:14_2} the so called two-stage dynamic stochastic model for description of  fusion-fission reaction with heavy ions was developed. In this model on the first stage of calculations the approaching of projectile nucleus to the target nucleus up to their touching point is considered. On the second stage we study the evolution of compact system, formed after merging of colliding nuclei.

In present work we apply the two-stage model \cite{GI:02,Litnevsky:14_1, Litnevsky:14_2} for the description of the kinetic energy and mass distributions of the fission fragments obtained in recent publication \cite{Andreev:18} for the reaction ${\rm ^{36}Ar + ^{142}Nd \to ^{178-x}Pt + xn}$ at beam energy $E_{\rm lab}= 153.9, 168.8$ and $178.8$ MeV.
This publication contains experimental data on the energy- and mass-distributions of fission fragments, average energy taken away by the pre-fission neutrons, average induced angular momentum and the rotational energy. Comparison of such data with calculated results would be a good test for the theoretical models.
%%%%%%%%%%%%%%%%%%%%%%%%%%%%%%%%%%%%%%%%%%%%%%%%%%%%%%%%%%%%%%%%%%%%%%%%%%%%%%%%%%%%%%
\section{The two-stage approach}
\begin{figure}[htp]
\includegraphics[width=0.4\textwidth]{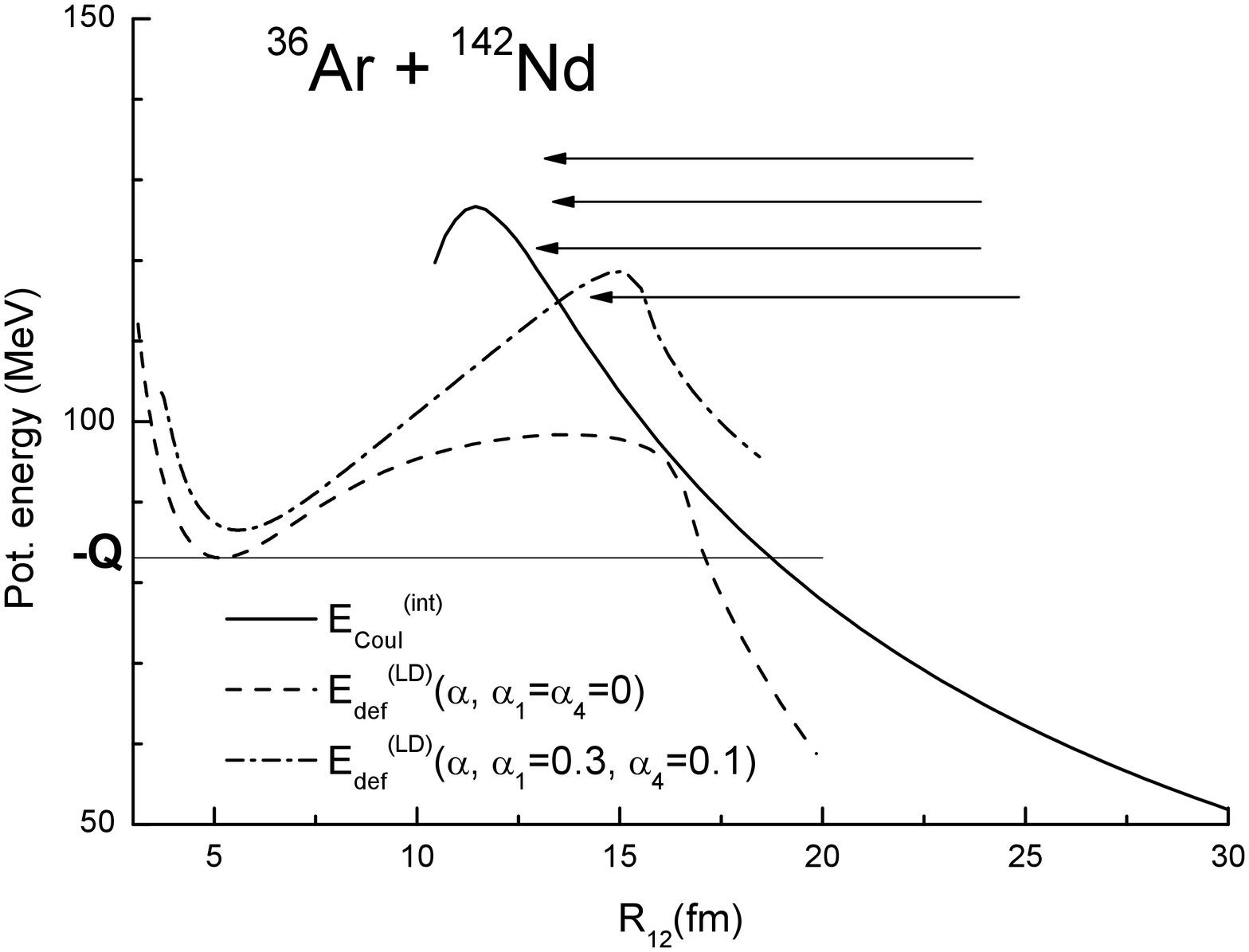}
\caption{The potential energy $V^{I}_{pot}$ \protect\req{eq:potI} of colliding ions ${\rm ^{36}Ar + ^{142}Nd}$ in the entrance channel (solid) and the liquid drop deformation energy of ${\rm ^{178}Pt}$ for mass-symmetric ($\alpha_3=\alpha_4=0$, dash) and mass-asymmetric ($\alpha_3=0.3, \alpha_4=0.1$, dot-dash) shapes as function of the distance $r$ between centers of mass. Horizontal line is the $Q$-value of reaction ${\rm ^{36}Ar + ^{142}Nd}{\longrightarrow\rm ^{178}Pt}$.}
\label{scission}
\end{figure}

%%%%%%%%%%%%%%%%%%%%%%%%%%%%%%%%%%%%%%%%%%%%%%%%%%%%%%%%%%%%%%%%%%%%%%%%%%%%%%%%%%%%%%
For the calculation of  fusion-fission reaction with heavy ions we use the two-stage dynamic stochastic model \cite{GI:02,Litnevsky:14_1, Litnevsky:14_2}. As it was mentioned above, on first stage of calculations we consider the   approaching of projectile nucleus to the target nucleus up to their touching point. On the second stage we study the evolution of compact system, formed after merging of colliding nuclei. The schematic comparison of the potential energy in fusion and fission channels is shown in Fig.~\ref{scission}.
The final point of the second stage calculations could be the splitting of the system back into two fragments or formation the evaporation residue $\rm ^{178-x}Pt$, where $x$ is a number of neutrons, evaporated by the compound system in order to reduce its excitation energy. Of course, the outcome depends on the reaction energy.

The evolution of both separated ions and compact system is described by Langevin equations for the collective parameters that fix the shape of the system.  For the approaching process the collective parameters are the parameters $\alpha_p$ and $\alpha_t$ of quadrupole deformation for both ions and the distance $r$ between their centers of mass.  On the approaching stage we also take into account the orientation of target nucleus -- the angle $\theta_t$ between the symmetry axis of the deformed in the ground state target nucleus and the line connecting centres of mass of colliding nuclei. The shape of the compact system is described by the parameters  $\alpha$, $\alpha_1$ and $\alpha_4$ of Cassini parameterization \cite{Pashkevich:71} that specify the total elongation, the mass asymmetry and neck radius of the system.
%One can clarify the role of each parameter looking at the Fig.~\ref{4param}.

\begin{figure}[htp]
\includegraphics[width=0.4\textwidth]{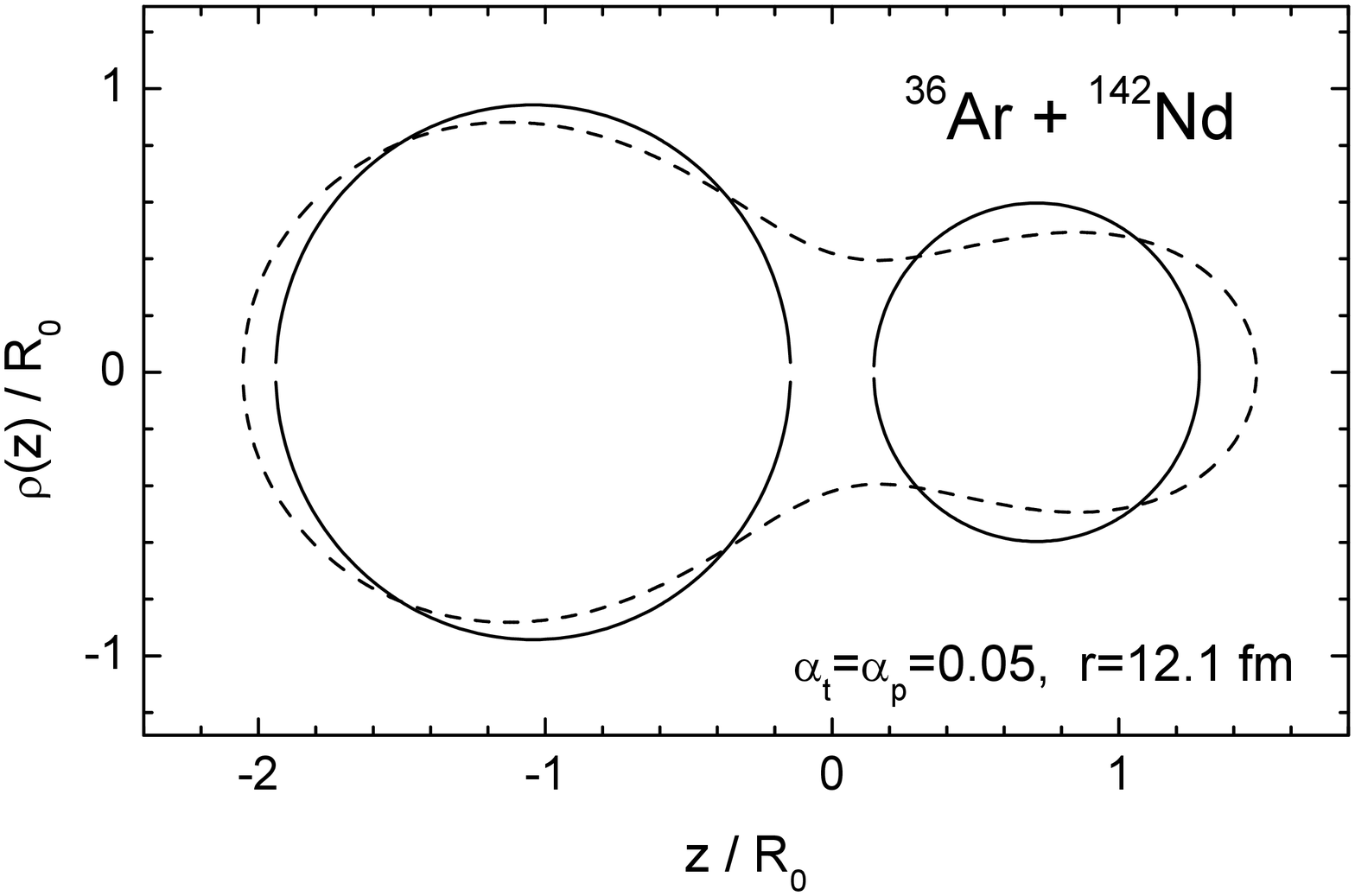}
\caption{An example of the touching configuration of ${\rm ^{36}Ar}$ and ${\rm ^{142}Nd}$ ($r$= 12.1 fm, $\alpha_t=\alpha_p=0.05$, solid) and the fit by 3-dimensional ($\alpha, \alpha_1, \alpha_4$) Cassini ovaloids (dash). }
\label{touching}
\end{figure}
At the end of the first stage calculations (at the touching point) we fix the center-of-mass distance, the potential and internal (dissipated) energy of the system.
At the initial point of the second stage calculations we use this data to define the shape of the system, formed after collision of the initial nuclei. To do it we use three conditions. First  we fix $\alpha_4$ and define $\alpha$ and $\alpha_1$ from the requirement that the center of mass distance $r$ and the mass asymmetry of separated ions and the compact system are the same. Then, we vary $\alpha_4$ and among all possible  shapes, which satisfies that requirement, we chose the one, that corresponds to the lowest potential energy $E_{\rm def}(\alpha, \alpha_1, \alpha_4)$ of compact shape.
%In this way we define the compact shape that corresponds to the same elongation and mass asymmetry as the separated ions at the touching point and to the lowest potential energy $E_{\rm def}(\alpha, \alpha_1, \alpha_4)$
An example of relation between the shapes of separated ions and the compact system is shown in Fig.~\ref{touching}.

At both stages of calculation the time evolution of the collective degrees of freedom   ${\bf q}\equiv\{\alpha, \alpha_n\}$ and corresponding momenta $ {\bf p}/{\bf m}\equiv\{\dot{\alpha}, \dot{\alpha}_n\}$ is described in terms of the Langevin equations
\cite{Abe:96,Marten:92}, namely,
\begin{eqnarray}
\dot {q}_\beta&=&\mu_{\beta \nu}p_\nu , \nonumber\\
\dot {p}_\beta&=&-\frac{1}{2}p_\nu p_\eta\frac{\partial \mu_{\nu \eta}}{\partial q_\beta}+K_\beta-\gamma_{\beta \nu}\mu_{\nu \eta}p_\eta+\theta _{\beta \nu}\xi_\nu.
\label{eq:UL}
\end{eqnarray}
Here ${q}_\beta$ are the deformation parameters and a convention of summation
over repeated indices $\nu$, $\eta$ is used.
The quantity $\gamma_{\beta\nu}$ is the tensor of friction coefficients and
$\mu_{\beta\nu}$ is the tensor inverse to the mass tensor
$m_{\beta\nu}$, $K_\beta$ is a component of the conservative force $\vec K=-\bigtriangledown F$, where $F=V_{pot}-aT^2$ is the free energy of the system, $V_{pot}$ -- its potential (deformation) energy, $a$ is the level density parameter \cite{mebel:92} and the temperature $T$ of system  is related to the internal (dissipated) energy by
the Fermi-gas formula $T=\sqrt {E_{\rm dis}/a}$.

Friction provides the dissipation of collective motion energy into internal energy. The fluctuations in the system are described  by the random force $\theta_{\beta\nu}\xi_\nu$.
Here $\xi_\nu$ is a random number with the following properties
\begin{eqnarray}
<\xi_\nu>&=&0,\nonumber\\
<\xi_\beta(t_1)\xi_\nu(t_2)>&=&2\delta_{\beta\nu}\delta(t_1-t_2).
\end{eqnarray}

The magnitude of the random force $\theta_{\beta\nu}$ is
expressed in terms of diffusion tensor
$D_{\beta\nu}=\theta_{\beta\eta}\theta_{\eta\nu}$, which is
related to the friction tensor $\gamma_{\beta\nu}$ via the modified
Einstein relation $D_{\beta\nu}=T^*\gamma_{\beta\nu}$, where $T^*$ is the effective temperature \cite{hofkid}.

The internal energy of the system could be calculated from the energy conservation condition:
\begin{eqnarray}
E_{\rm cm}=V_{\rm pot}+E_{\rm kin}+E_{\rm dis}, \label{eq:energ_conserv_low}
\end{eqnarray}
here $E_{\rm cm}=E_{\rm lab}A_{\rm Nd}/(A_{\rm Nd}+A_{\rm Ar})$ is the reaction energy, calculated in the center-of-mass system, and $E_{\rm kin}$ is the kinetic energy of the collective motion.

Some terms of the equation (\ref{eq:UL}) should be determined twice, ones for the first, and ones for the second stage of calculations. Such terms we will denote by the upper indexes  ($I$) and ($II$), respectively.

The deformation energy $E_{\rm def}^{(t)}$ and $E_{\rm def}^{(p)}$ of colliding ions
and $E_{\rm def}$ of the combined system are calculated within the macroscopic-microscopic shell-correction approach proposed by Strutinsky \cite{Strutinsky:67,Brack:72,Pashkevich:71}.
The interaction potential between the colliding ions consists of Coulomb part $V_{\rm Coul}$ \cite{kurmanov} and nuclear $V_{\rm GK}$ -- Gross-Kalinowski potential \cite{Gross:78} , which was modified \cite{Marten:92} in order to describe the interaction of deformed ions.
The rotation of the system is also taken into account on the both stages of calculations:
\bel{erot}
E^{I}_{\rm rot}=\frac{\hslash^2 L(L+1)}{2(Mr^2+J_t+J_p)}, \qquad E^{II}_{\rm rot}=\frac{\hslash^2 L(L+1)}{2J},
\end{equation}
where
$M=M_tM_p/(M_t+M_p)$ is the reduced mass of target and projectile,
$J_t$, $J_p$ and $J$ are the rigid body moments of inertia of the target, projectile nucleus and of the combined system, respectively, and $L$ is an angular momentum of the whole system.

Finally, the potential energy of the system is:
\begin{eqnarray}
V^{I}_{\rm pot}&=&V_{\rm Coul}+V_{\rm GK}+E_{\rm def}^{(t)}+E_{\rm def}^{(p)}+E^{I}_{\rm rot}, \label{eq:potI} \\
V^{II}_{\rm pot}&=&E_{\rm def}+E^{II}_{\rm rot}. \label{eq:potII}
\end{eqnarray}
Besides the deformation energy, the dynamic properties of
each nucleus are characterized by the friction $\gamma_{\beta\nu}$
and inertia $m_{\beta\nu}$ tensors, that were calculated within the linear response
approach and local harmonic approximation \cite{Hofmann:97,Hofmann:08}. In
this approach many quantum effects such as shell and pairing
effects, and the dependence of the collisional width of single particle
states on the excitation energy, are taken into account.
The precise expressions for the components of the friction $\gamma_{\beta\nu}$ and inertia $m_{\beta\nu}$ tensors can be found in \cite{Ivanyuk:1999}.
Tensors $\gamma_{\beta\nu}$ and $m_{\beta\nu}$ \cite{Ivanyuk:1999} characterize completely the inertia and friction properties of the combined system and were used in our previous calculations within three-dimensional Langevin approach with Cassini shape parameterization and the two-center shell model shape parameterization \cite{usang2016}.

For the first stage of calculation, besides the internal processes in ions, one should account also for their relative motion.
The inertial tensor $m^{I}_{\beta\nu}$ in this case has only diagonal components: $m^{I}_{\rm rr}=M$ (reduced mass), $m^{I}_{\alpha_{\rm t}\alpha_{\rm t} }=m^{\rm t}_{\alpha \alpha}$, $m^{I}_{\alpha_{\rm p}\alpha_{\rm p} }=m^{\rm p}_{\alpha \alpha}$ and $m^{I}_{\theta_{\rm t}\theta_{\rm t} }=J_{\rm t}$.
The friction tensor for the first stage of the process was defined as the sum of friction tensor for the relative motion , obtained in the surface-friction model \cite{Frobrich:84}, like it was done in \cite{Litnevsky:75:2012}, and the diagonal friction tensor for colliding ions, which has only two non zero components ($\gamma^{I}_{\alpha_{\rm t}\alpha_{\rm t} }=\gamma^{\rm t}_{\alpha \alpha}$; $\gamma^{I}_{\alpha_{\rm p}\alpha_{\rm p} }=\gamma^{\rm p}_{\alpha \alpha}$).
%%%%%%%%%%%%%%%%%%%%%%%%%%%%%%%%%%%%%%%%%%%%%%%%%%%%%%%%%%%%%%%%%%%%%%%%%%%%%%
\section{The results of first stage calculations}
\begin{figure}[htp]
\includegraphics[width=0.48\textwidth]{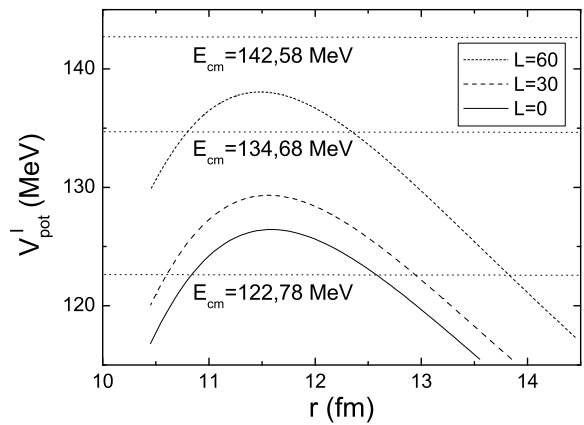}
\caption{Potential energy of the system as a function of the distance $r$ between target and projectile. The orientation of target $\theta_t=0$, deformation parameters $\alpha_t$ and $\alpha_p$ correspond to the ground state shapes of the target and projectile ions. }
\label{In_Pot}
\end{figure}

\begin{figure}[htp]
\includegraphics[width=0.48\textwidth]{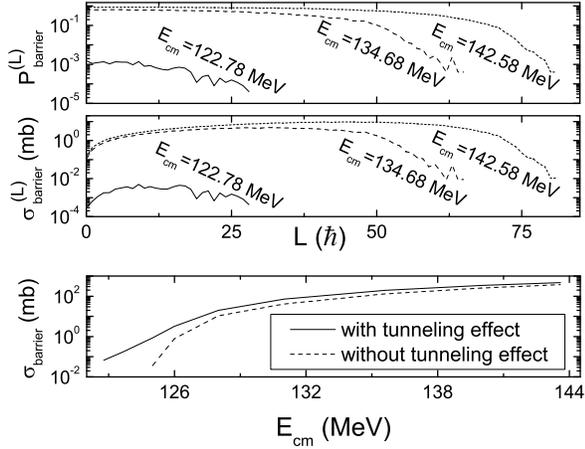}
\caption{Top: the dependence of the probability and partial cross section (middle) of the penetration through the potential barrier, calculated with account of the tunneling effect, on the angular momentum of the system. Bottom: the dependence of the cross section of the penetration through the potential barrier on the reaction energy. }
\label{Sig_bar}
\end{figure}

%%%%%%%%%%%%%%%%%%%%%%%%%%%%%%%%%%%%%%%%%%%%%%%%%%%%%%%%%%%%%%%%%%%%%%%%%%%%%%
In the present paper, the first stage calculations were ended as soon as the system overcomes the potential barrier. The position of the barrier depends on deformations and orientation of the colliding ions. It also slightly depends on angular momentum of the system (see fig.~\ref{In_Pot}).
%%%%%%%%%%%%%%%%%%%%%%%%%%%%%%%%%%%%%%%%%%%%%%%%%%%%%%%%%%%%%%%%%%%%%%%%%%%%%%%%%%%%%%%%%%%%%%
%%%%%%%%%%%%%%%%%%%%%%%%%%%%%%%%%%%%%%%%%%%%%%%%%%%%%%%%%%%%%%%%%%%%%%%%%%%%%%%%%%%%%%%%%%%%%%
As one can see, some experimental data in \cite{Andreev:18} are obtained for the sub-barrier energies. In order to describe such reactions we took into account the quantum tunneling through the potential barrier.
The penetrability of the barrier was defined in the WKB approximation \cite{Nevzorova:08} as
\begin{equation}
T_L(E)=\left[ 1+\exp\left( \frac{2}{\hslash} \int_{r_2}^{r_1}\sqrt{2m(V^{\rm fus}-E)}dr\right) \right]^{-1},
\label{eq:TL}
\end{equation}
where the integration is carried out between the turning points
$r_1$ and $r_2$ in the subbarrier region and $E$
is the incident energy,  equal to the
potential energy of the system at the turning points. As one can see from
Fig.~\ref{Sig_bar}  (bottom), the account of quantum tunneling increases substantially the probability of
penetration throw the barrier in the subbarrier region.

%%%%%%%%%%%%%%%%%%%%%%%%%%%%%%%%%%%%%%%%%%%%%%%%%%%%%%%%%%%%%%%%%%%%%%%%%%%%%%%%%%%%%%%%%%%%%%%%%%%%%%%%
%%%%%%%%%%%%%%%%%%%%%%%%%%%%%%%%%%%%%%%%%%%%%%%%%%%%%%%%%%%%%%%%%%%%%%%%%%%%%%%%%%%%%%%%%%%%%%%%%%%%%%%%
The angular momentum is a free parameter of our model. So, the first stage calculations are carried out for various values of $L$.
One can see from Fig.~\ref{In_Pot}, that the height of potential barrier increases with the angular momentum of the system. The probability to overcome the barrier will decrease in this case, see Fig.~\ref{Sig_bar}. Knowing  the probability of event, we can calculate its partial, Fig.~\ref{Sig_bar} ( middle) and total,  Fig.~\ref{Sig_bar} (bottom) cross sections,  respectively:

\begin{eqnarray}
\sigma^L_{\rm barrier}(L,E_{\rm cm})&=&\pi \rlap{\(\bar{\phantom{a}}\)}\lambda ^2
(2L + 1)P_{\rm barrier}(L,E_{\rm cm}),\qquad~
\label{eq:part_cross_section}\\
\sigma_{\rm barrier}(E_{\rm cm})&=&\sum_L \sigma^L_{\rm barrier}(L,E_{\rm cm}).
\label{eq:total_cross_section}
\end{eqnarray}

The main goal of the first stage calculations is to get the potential energy $V^{I}_{\rm pot}$ of the system, and its internal (dissipated) energy $E_{\rm dis}$ at the touching point.
\begin{figure}[htp]
\includegraphics[width=\columnwidth]{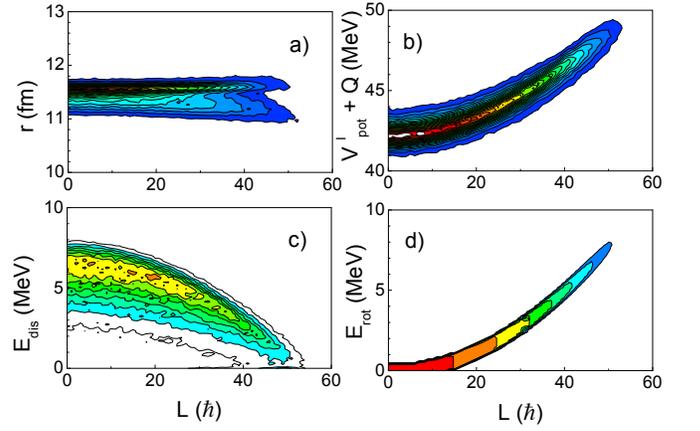}
\caption{The distributions of center-of-mass distance (a), potential (b), dissipated (c) and rotational (d) energies of the system as functions of angular momentum $L$ at the touching point for $E^*$= 50.5 MeV.}
\label{InData}
\end{figure}

Having at our disposal the distributions given in Fig.~\ref{Sig_bar} (middle) and Fig.~\ref{InData}b,c,d we can specify (by hit and miss method) the initial values of angular momentum, internal and potential energy of the system for the beginning of the second stage of calculations. The way to choose the initial shape parameters of the combined system was discussed in the previous section, see Fig.~\ref {touching} and the text around it. So, the initial values for the second stage calculations are defined by the final data from the first stage
\belar{initial}
E_{kin, in}^{II}(L)&=&E_{kin}^{I}(L), \nonumber\\
E_{dis, in}^{II}(L)&=&E_{dis}^{I}(L)+V_{pot}^I-V_{pot}^{II}.
\end{eqnarray}
In order to bring $V^{II}_{\rm pot}$ and $V^{I}_{\rm pot}$ to the same scale, the so called $Q$-value of reaction should be added to $V^{II}_{\rm pot}$
\bel{VIIin}
V^{II}_{\rm pot, in}\longrightarrow V^{II}_{\rm pot, in}+Q
\end{equation}
where $Q<0$ is defined in terms of ground state energies $Q\equiv E_{gs}^{(t+p)}
-E_{gs}^{(t)}-E_{gs}^{(p)}$. One can also define the excitation energy $E^*$ of reaction
\bel{Qreact}
E^*=E_{\rm cm}+Q.
%-E_{kin}^{I}(L)-E_{dis}^{I}(L)-E_{rot}^{I}(L).
\end{equation}
The distribution in Fig.~\ref{InData}d is the distribution of rotational energy at the touching point. The width of this distribution is very small. The contribution to this width comes from the uncertainly of center-of-mass distance $r$ (which is also small, see Fig.~\ref{InData}a) that appears in the moment of inertia in Eq.~\req{erot}.

Each point in Fig.~\ref{InData}d is the contribution of trajectory $"i"$ that reached the touching point. By summing over all trajectories one can define the average angular momentum $\langle L \rangle$ and the average rotational energy $\langle E_{rot} \rangle$ of the system at the touching point,
\bel{Lavr}
\langle L \rangle = \sum_i L_i \slash \sum_i 1\,,\quad\langle E_{rot} \rangle = \sum_i E_{rot}^i \slash \sum_i 1\,.
\end{equation}
The summation over trajectories based on distribution of Fig.~\ref{InData}d leads to the value $\langle L \rangle\approx20.5~ \hbar$ and $\langle E_{rot} \rangle\approx$1.89 MeV. In the same way we can calculate the average value of dissipated energy $\langle E_{dis} \rangle\approx$ 4.40 MeV.
As we can see, the sum of rotational and dissipated energy at the touching point is not so large compared with the excitation energy $E^*$=50.5 MeV.
%%%%%%%%%%%%%%%%%%%%%%%%%%%%%%%%%%%%%%%%%%%%%%%%%%%%%%%%%%%%%%%%%%%%
\section{The evolution of compact system}
\begin{figure}[htp]
\includegraphics[width=0.48\textwidth]{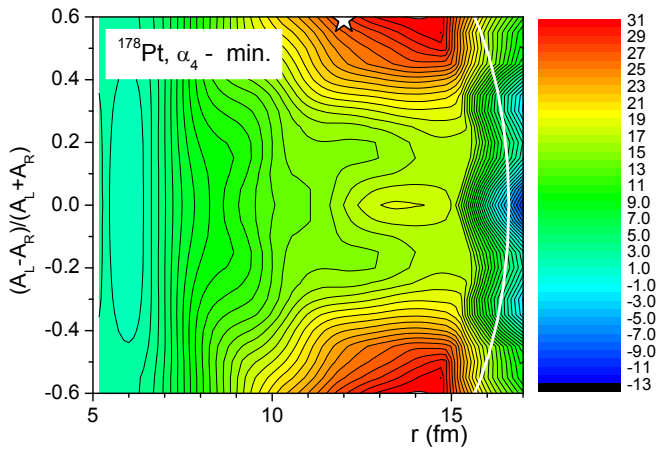}
\caption{The minimized with respect to $\alpha_4$ deformation energy of $^{178}$Pt at $T$=0. The white line mark the position of zero neck for $\alpha_4=0$.}
\label{Edef178}
\end{figure}

%%%%%%%%%%%%%%%%%%%%%%%%%%%%%%%%%%%%%%%%%%%%%%%%%%%%%%%%%%%%%%%%%%%%
The potential energy surface of $^{178}$Pt in coordinates $r$ - mass asymmetry, $(A_L-A_R)/(A_L+A_R)$, is shown in Fig.~\ref{Edef178}.  The star in this figure marks the position of initial point corresponding to the shape shown in Fig.\ref{touching}.

Starting from the touching point under the action of random force the system would move in the direction of ground state, form the compound nucleus and then undergo fission.
The evolution of the compound nucleus could last long enough to evaporate gammas or  light particles, so   we should take into account this de-excitation process.
The evaporation of particles and $\gamma$-quanta by the excited
compact system is described in our approach within the statistical model
\cite{mebel:92}. On each integration step the partial
width of the corresponding decay channel \cite{mebel:92} is calculated, then by the
hit-and-miss method we decide whether some particle was emitted and which
kind of particle was emitted. If some particle was emitted, the
binding energy of this particle was subtracted from the
excitation energy of the system. Also the particle can carry away some energy (its kinetic energy).
\begin{figure}[t]
\includegraphics[width=0.48\textwidth]{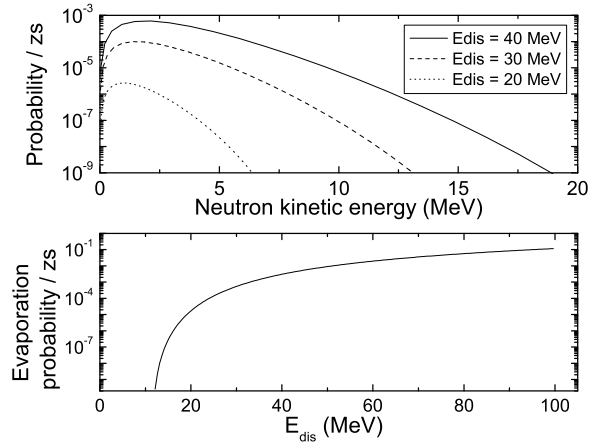}
\caption{The probability (per zepto second) to evaporate the neutron with some value of the kinetic energy (top) and the neutron evaporation probability (per zepto second) as function of nuclear internal energy (bottom).}
\label{neutrons}
\end{figure}

The main evaporation channel is the evaporation of neutrons. Probability of the neutron evaporation and its kinetic energy depends on the dissipated energy of the system  (see Fig.~\ref {neutrons}).
One can see, that system with dissipated energy $E_{\rm dis}=40$ MeV, for example, should live on average $400$ zs before first neutron is evaporated.
\begin{figure}[htp]
\includegraphics[width=0.40\textwidth]{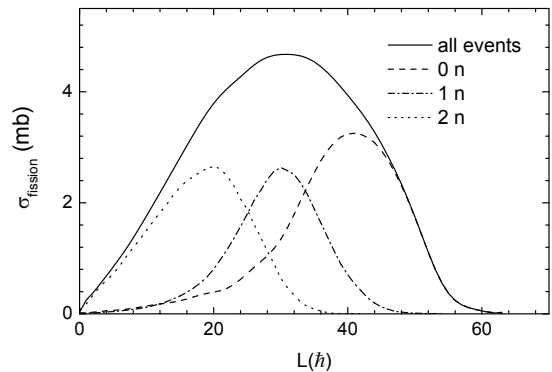}
\caption{The partial cross sections of the fission process in case of zero (dash), one (dash-dot) or two (doted curve) neutrons emitted by compound nucleus before fission ias function of angular momentum of the system. Solid curve is the partial cross sections of all fission events.}
\label{fission}
\end{figure}
After evaporation it loses about $10$ MeV as a neutron kinetic and binding energy, so, to evaporate the second neutron it needs about $1500$ zs, and so on. Of course, during this evolution time there is a competition between neutron evaporation and fission. The fission of  nucleus can occur before it  evaporates any neutron (dashed line), or it could evaporate one (short dashed line), two (dotted line) or more neutrons before fission (see Fig. \ref{fission}). The maximal number of the evaporated neutrons depends on the initial excitation energy of the system $E^*$, see Eq.~\req{Qreact}.

\begin{figure}[htp]
\includegraphics[width=0.45\textwidth]{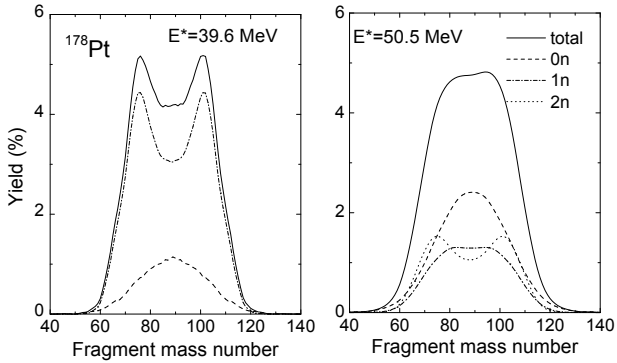}
\caption{The contributions to the total  fission cross section of $^{278}$Pt (solid) at $E^*$=39.6 MeV and 50.5 MeV of the fission events without neutron evaporation (dash), with evaporation of one neutron (dash-dot) and two (dot) neutrons. }
\label{MassDistr}
\end{figure}

After evaporation of several neutrons the systems internal energy decreases and the role of the nuclear shell structure become more and more significant.
In Fig.~\ref{MassDistr} one can see the evolution of fragment mass distribution due to the de-excitation  process (for the initial excitation energies $E^*=39.6$ MeV and $E^*=50.5$ MeV). The mass distribution is symmetric if the system undergoes fission before it evaporates any neutron. Then, after the first neutron evaporation, the  system slightly cools down and the influence of the shell effects will be noticeable. And, finally, after second neutron evaporation the distribution becomes strongly asymmetric.
%%%%%%%%%%%%%%%%%%%%%%%%%%%%%%%%%%%%%%%%%%%%%%%%%%%%%%%%%%%%%%%%%%%%%%%%%%
%%%%%%%%%%%%%%%%%%%%%%%%%%%%%%%%%%%%%%%%%%%%%%%%%%%%%%%%%%%%%%%%%%%%%%%%%%%

\begin{figure}[htp]
\includegraphics[width=0.48\textwidth]{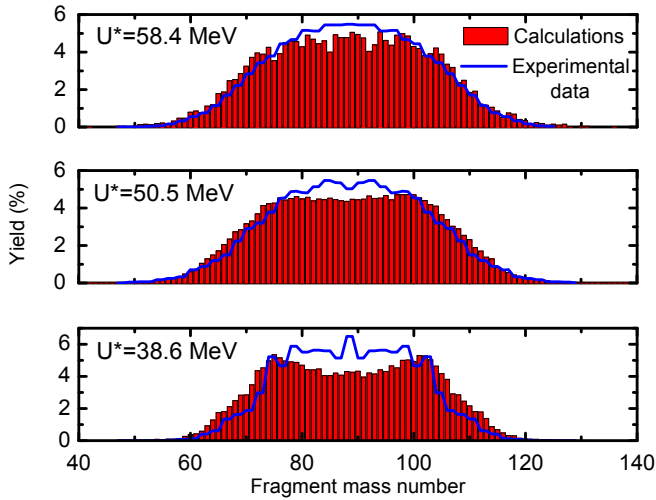}
\caption{The calculated fission fragment mass distributions in comparison with the experimental data \cite{Andreev:18}.}
\label{TotalMassDistr}
\end{figure}

The total mass distribution obtained by superimposing of distributions shown in Fig.~\ref{MassDistr} is compared in Fig.~\ref{TotalMassDistr} with the experimental data \cite{Andreev:18}. One can see that the calculated mass distributions for all three values of excitation energy $E^*$ are very close to the experimental data.
%%%%%%%%%%%%%%%%%%%%%%%%%%%%%%%%%%%%%%%%%%%%%%%%%%%%%%%%%%%%%%%%%%%%%%%%%%%
%%%%%%%%%%%%%%%%%%%%%%%%%%%%%%%%%%%%%%%%%%%%%%%%%%%%%%%%%%%%%%%%%%%%%%%%%
\section{The total kinetic energy}
\begin{figure}[h]
\includegraphics[width=\columnwidth]{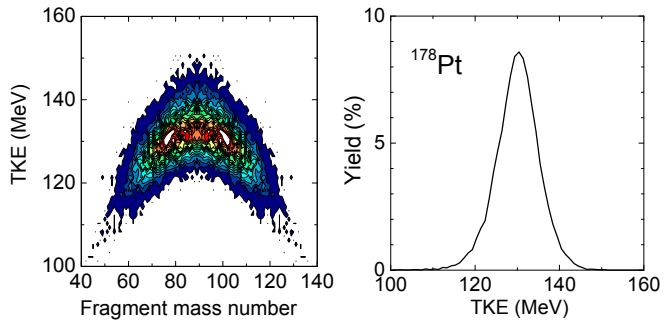}
\caption{Left: The mass-energy distribution of fission events for $E^*$=50.5 MeV. Right: The normalized to 100 \% yield of fission fragments as function of the total kinetic energy. }
\label{fig_tke}
\end{figure}

%%%%%%%%%%%%%%%%%%%%%%%%%%%%%%%%%%%%%%%%%%%%%%%%%%%%%%%%%%%%%%%%%%%%%%%
The mass-energy distribution of fission events is shown in the left part of Fig.~\ref{fig_tke}. In these
calculations for each trajectory that reached the scission point $\{q_i\}$ we define $TKE_i$ as the sum of the Coulomb interaction
of spherical fragments at the scission point and
prescission kinetic energy $KE^{(pre)}$,
\bel{tke}
TKE_i=E_{Coul}^{(int)}(q_i) + KE_i^{(pre)}.
\end{equation}
Here
\bel{Ecoul}
E_{Coul}^{(int)}(q_i) \equiv Z_LZ_H e^2/r(q_i),
\end{equation}
%%%%%%%%%%%%%%%%%%%%%%%%%%%%%%%%%%%%%%%%%%%%%%%%%%%%%%%%%%%%%%%%%%%%%%%%%%%%%%%%%%%%%%%%%%
%%%%%%%%%%%%%%%%%%%%%%%%%%%%%%%%%%%%%%%%%%%%%%%%%%%%%%%%%%%%%%%%%%%%%%%%%%%%%%%%%%%%%%%%%%%
where $eZ_L $ and $eZ_H $ are the charges of light and heavy fragments. The prescission kinetic energy $KE_i^{(pre)}$ is the kinetic energy in fission direction obtained from the solutions of Langevin equations at the scission point. It turned out in these calculations that the average value of $KE^{(pre)}$ is very small, of the order of $1\div 2$ MeV and the main contribution to TKE comes from the Coulomb repulsion energy.

The TKE-distribution of fission fragments is shown in the right part of Fig.~\ref{fig_tke}. The shape of distribution is very close to a single Gaussian.
%%%%%%%%%%%%%%%%%%%%%%%%%%%%%%%%%%%%%%%%%%%%%%%%%%%%%%%%%%%%%%%%%%%%%%%%%%%%%%%%%%%%%%%%%%
\begin{figure}[h]
\includegraphics[width=0.8\columnwidth]{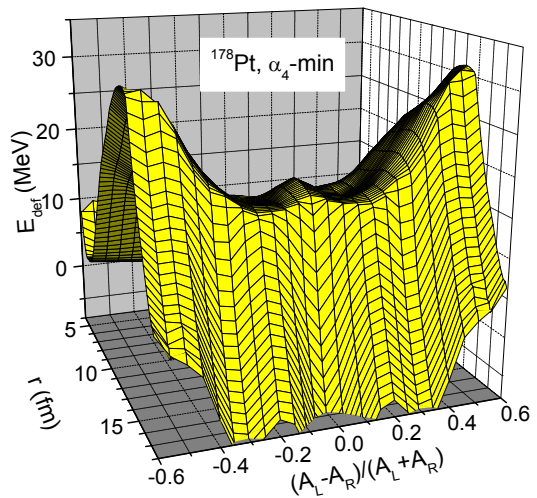}
\caption{The "backside" of potential energy surface: the saddle-to-scission part of deformation energy of $^{278}$Pt minimized with respect to $\alpha_4$ at $T=0$. }
\label{vallies}
\end{figure}
%%%%%%%%%%%%%%%%%%%%%%%%%%%%%%%%%%%%%%%%%%%%%%%%%%%%%%%%%%%%%%%%%%%%%%%%%%%%%%%%%%%%%%%%%%%
We did not find the contributions to TKE-distribution from two fission modes as it was argued in \cite{Andreev:18}. We see the two fission valleys on potential energy surface at large elongation, beyond the saddle point, see Fig.~\ref{vallies}. Between saddle and scission the mass-symmetric valley is even deeper compared with mass-asymmetric. However, it follows from the calculations that the dynamical trajectories do not follow the bottom of fission valley. The mass- and energy-distributions are formed mainly at the saddle. The trajectories spend a lot of time inside saddle. Some of them reach the saddle and move very quickly towards the scission. The descent from saddle to scission in $^{178}$At (and other light fissioning nuclei) is very short, see Fig.~\ref{vallies}, and steep. The trajectories do not have much time to adjust themselves to the potential energy landscape during  saddle-to-scission motion. Note also, that the mass asymmetry of asymmetric valley beyond the saddle is much larger as (coinciding with experimental) mass asymmetry of PES at the saddle.

The average value of TKE is equal to 130.4 Mev what is rather close to the position of main peak, TKE$^{\rm high}$=133.4 MeV, in experimental results \cite{Andreev:18}. The width of TKE-distribution $\sigma_{TKE}$= 11 MeV is, however, much smaller than experimental value, see Fig.~2a of \cite{Andreev:18}. The difference might be attributed to the uncertainty of scission point configuration formed in the reaction, the finite (in)accuracy of TKE measurements and some limitations of theoretical approach.
%In particular, the generalization of tree-dimensional Langevin approach to a four-dimensional would increase the variety of shapes at the scission point and, thus, could increase the width of TKE-distribution.
\section{Acknowledgments}
This study includes the results of "Research and development of an innovative transmutation system of LLFP by fast reactors", entrusted to the Tokyo Institute of Technology by the Ministry of Education, Culture, Sports, Science and Technology of Japan (MEXT). The authors would like to thank to Prof. A.N. Andreyev and Prof. I.~Tsekhanovich for useful discussions.
%%%%%%%%%%%%%%%%%%%%%%%%%%%%%%%%%%%%%%%%%%%%%%%%%%%%%%%%%%%%%%%%%%%%%%%%
\section{Conclusions}
%%%%%%%%%%%%%%%%%%%%%%%%%%%%%%%%%%%%%%%%%%%%%%%%%%%%%%%%%%%%%%%%%%%%%%%%
Within the developed earlier two-stage approach for the fusion-fission reactions we have studied the reaction ${\rm ^{36}Ar + ^{142}Nd \to ^{178-x}Pt + xn}$. The obtained results are compared with the available experimental data. The calculated mass distributions of fission fragments for
three energies $E_{cm}$=122.78, 134.68 and 142.58 MeV are in good agreement with the experimental data \cite{Andreev:18}.
 The most probable mass division for asymmetric component of fission fragments mass distribution (Fig.~\ref{TotalMassDistr}) is found to be about $A_{\rm L}/A_{\rm H}=79/98$, what is rather close to the experimental data \cite{Andreev:18}. From our calculations it follows that the competition between symmetric and asymmetric fission channels is caused by the enhancement of shell effects in  the compound nucleus due to the process of its de-excitation by neutron emission.

 The average value of TKE is equal to 130.4 Mev what is rather close to the position of main peak, TKE$^{\rm high}$=133.4 MeV, in experimental results. The calculated shape of distribution is very close to a single Gaussian. We did not find the contributions to TKE-distribution from two fission modes as it was argued in \cite{Andreev:18}.

From the results of calculations it follows that the dynamical trajectories do not follow the bottom of fission valley and the mass- and energy-distributions of fission fragments of $^{178}$Pt are formed mainly at the saddle.
%%%%%%%%%%%%%%%%%%%%%%%%%%%%%%%%%%%%%%%%%%%%%%%%%%%%%%%%%%%%%%%%%%%%%%%%%%%%%%%%%%%%%%%%%%%

\end{document}